\documentclass[a4paper,11pt]{article}
\pdfoutput=1 

\usepackage{jcappub} 

\usepackage[T1]{fontenc} 

\def\be{\begin{equation}}
\def\ee{\end{equation}}
\def\beq{\begin{eqnarray}}
\def\eeq{\end{eqnarray}}

\title{Sudden singularities in generalized hybrid metric-Palatini cosmologies}

\author[a,1]{João Luís Rosa,\note{Corresponding author.}}
\author[b,c]{Francisco S. N. Lobo,}
\author[d]{Diego Rubiera-Garcia}

\affiliation[a]{Institute of Physics, University of Tartu, W. Ostwaldi 1, 50411 Tartu, Estonia}
\affiliation[b]{Instituto de Astrof\'{\i}sica e Ci\^{e}ncias do Espa\c{c}o, Faculdade de
Ci\^encias da Universidade de Lisboa, Edif\'{\i}cio C8, Campo Grande,
P-1749-016 Lisbon, Portugal}
\affiliation[c]{Departamento de F\'{\i}sica, Faculdade de
Ci\^encias da Universidade de Lisboa, Edif\'{\i}cio C8, Campo Grande,
P-1749-016 Lisbon, Portugal}
\affiliation[d]{Departamento de F\'isica Te\'orica and IPARCOS, Universidad Complutense de Madrid, E-28040
Madrid, Spain}

\emailAdd{joaoluis92@gmail.com}
\emailAdd{fslobo@fc.ul.pt}
\emailAdd{drubiera@ucm.es}

\abstract{In this work, we explore cosmological sudden singularities arising in the dynamically equivalent scalar-tensor representation of generalized hybrid metric-Palatini gravity. Using a FLRW background, we show that the structure of the field equations prevents sudden singularities from arising at time derivatives of the scale factor of orders lower than four, but that they are allowed to appear for time derivatives of higher orders. Imposing an ansatz for the scale factor, we provide an explicit solution where these sudden singularities appear in the fourth-order time derivative of the scale factor. A comparison of the Hubble and deceleration parameters arising from this model with the experimental measurements from the Planck Satellite allow us to impose constraints on the time span for which the occurrence of sudden singularities becomes likely in our universe, as measured from the Big Bang.}

\begin{document}
\maketitle
\flushbottom

\section{Introduction}\label{sec:intro}

A central theme in modern cosmology is that our Universe is currently undergoing an era of acceleration expansion \cite{Perlmutter:1998np,Riess:1998cb}. Most of the theoretical models assume that the cause of this late-time cosmic speed-up is due to the presence of an exotic fluid with a negative pressure, dubbed as dark energy \cite{Copeland:2006wr}. In fact, according to the Planck Satellite 2018 results \cite{Aghanim:2018eyx}, the equation of state of this dark component is given by $\omega=-1.03 \pm 0.03$, and is therefore still compatible with a cosmological constant $\Lambda$. Now, equations of state with $\omega \leq -1$ correspond to phantom fields violating the null energy condition (NEC) \cite{Caldwell:1997ii}. This feature implies that the background cosmological evolution may undergone a singular behaviour in some cosmological parameters, at a given finite future time $t_s$. Such an event is dubbed as a {\it future cosmological singularity}, which involves the divergence of the scale factor and its derivatives, and include the Big Rip \cite{Caldwell:2003vq}, in which the fact the scale factor goes to infinity at $t=t_s$ (the Rip time). This scenario implies the impossibility of geodesic completeness  \cite{Ellis:1977pj} and prevents the survival any structure in the universe.

These future cosmological singularities obviously pose a threat to the predictability and determinism of our physical theories, and therefore much effort has recently been invested in order to systematically classify them \cite{Nojiri:2005sx,Frampton:2011sp,Frampton:2011rh,Dabrowski:2009kg,BeltranJimenez:2016dfc}. These classification schemes of future singularities are based on the cosmological parameters having a divergent behaviour, which typically are the scalar factor, the Hubble factor and its higher derivatives, as well as the energy density and pressure of the matter fields. A particularly interesting case, that will be analysed in detail throughout this work, is that of a sudden singularity, which is defined as a future event in the cosmological evolution in which the scale factor, the expansion rate, and the energy density are all finite, but the pressure diverges, thus inducing a divergence in higher-order derivatives of the scale factor.
In addition to determining the geodesically (in)complete character of a given spacetime containing a future cosmological singularity, one is also interested in its strength, i.e., whether it destroys bound structures or not, such as according to the strong/weak criteria of singularities introduced by Tipler \cite{Tipler:1977zza} and Clarke and Krolak \cite{CK,Krolak}, which has also been systematically applied in the cosmological context \cite{FernandezJambrina:2006hj}.

The occurrence of a future singularity is essentially determined by the assumptions of a specific cosmological model, such as the matter content and the theory of gravity supporting the model. Given such a model, current cosmological observations may allow, in principle, to constrain the time in which any such future singularity will occur \cite{Jimenez:2016sgs}. In this context, scalar-tensor theories of gravity and their cosmological consequences have been widely studied in the literature (for some reviews see \cite{Clifton:2011jh,Nojiri:2017ncd,Saridakis:2021lqd}), being constrained by gravitational wave data \cite{Langlois:2017dyl,Langlois:2018dxi} and cosmological observations \cite{Alonso:2016suf}. In particular, in Ref. \cite{Barrow:2019cuv} it was shown that sudden singularities (originally introduced in \cite{Barrow:2004xh} and further explored in \cite{Barrow:2004hk,Barrow:2004he,Lake:2004fu,Dabrowski:2005fg,Barrow:2008sn,Barrow:2009df,Barrow:2010ij,Barrow:2010wh,Barrow:2011ub,Barrow:2013tf,Perivolaropoulos:2016nhp,Barrow:2020rhh}) that respect the energy conditions can occur at finite times in Brans-Dicke and more general scalar-tensor theories of gravity.
These solutions were constructed explicitly in the Friedmann universes. While geodesic completeness remains valid \cite{Barrow:2013ria} and physical observers are not necessarily crushed (weak singularity), other pathologies potentially arise \cite{Barrow:2020ekb}. It was also shown in \cite{Barrow:2019cuv} that higher-order versions of these singularities, which involve singularities in derivatives of the scale factor that are higher than second order, are also possible, including those that arise when scalar fields have a self-interaction potential of a power-law form.

The main aim of this work is to study the existence of sudden singularities in a generalized hybrid metric-Palatini gravity theory, an extension of the $f\left(R\right)$ theories of gravity where the action depends not only in the Ricci scalar $R$ of the metric but also in a Palatini Ricci scalar built in terms of an independent connection \cite{Tamanini:2013ltp}. In $f\left(R\right)$ gravity, it is known that the metric and the Palatini approaches give rise to two distinct theories \cite{Sotiriou:2008rp}, which account for the late time cosmic acceleration of the universe without the need for dark energy sources. In fact, the physical motivation behind a hybrid combination, containing elements from both the metric and Palatini formalisms, turns out to be very successful in accounting for the observed phenomenology, such as, it is able to unify the late-time cosmic acceleration with the solar system constraints, and avoids some drawbacks of the original approaches. 
Indeed, an inherent drawback of the metric $f\left(R\right)$ is that it requires the chameleon mechanisms to correctly satisfy the solar-system constraints \cite{Khoury:2003aq,Khoury:2003rn}, whereas the Palatini $f\left(R\right)$ features microscopic matter instabilities, and is unable to describe the evolution of cosmic perturbations, and might induce surface singularities in polytropic stellar models, among other problems \cite{Olmo:2011uz, Gomez:2020rnq}. These problems are avoided in the hybrid combination.

The generalized hybrid metric-Palatini gravity theory can be written as a scalar-tensor one with two fields with their corresponding kinetic terms and an interaction potential \cite{Rosa:2017jld,Rosa:2018jwp}. 
In fact, generalized hybrid metric-Palatini gravity extends the original linear version proposed in \cite{Harko:2011nh}, where it was shown that the theory passes the Solar System observational constraints even if the scalar field is very light. This implies the existence of a long-range scalar field, which is able to modify the cosmological \cite{Capozziello:2012ny,Carloni:2015bua} and galactic dynamics \cite{Capozziello:2012qt,Capozziello:2013yha}, but leaves the Solar System unaffected \cite{Capozziello:2013uya,Capozziello:2013wq,Dyadina:2019dsu}. Further cosmological scenarios \cite{Boehmer:2013oxa,Santos:2016tds,Kausar:2019iwu,Sa:2020qfd,Sa:2020fvn,Paliathanasis:2020fyp,rosadynsys}, observational constraints \cite{Lima:2014aza,Lima:2015nma,Leanizbarrutia:2017xyd,Avdeev:2020jqo,Rosa:2021lhc}, stability issues \cite{Koivisto:2013kwa,Capozziello:2013gza,Chen:2020evr}, gravitational waves \cite{Kausar:2018ipo,Bombacigno:2019did}, higher dimensional \cite{Fu:2016szo,Rosa:2020uli}  and stringlike solutions \cite{Harko:2020oxq,Da:2021pbj} were explored. Solutions of compact objects such as black holes \cite{Danila:2018xya,Rosa:2020uoi,Bronnikov:2019ugl,Bronnikov:2020vgg}, stars \cite{Danila:2016lqx} and wormholes geometries \cite{Capozziello:2012hr,KordZangeneh:2020ixt} were also extensively analysed. We refer the reader to Refs. \cite{Capozziello:2015lza,Harko:2018ayt,Harko:2020ibn} for reviews on the hybrid metric-Palatini gravitational theory.

Here, we show that the structure of the field equations of these theories prevents such sudden singularities to arise at orders lower than four in the time derivative of the scale factor. For higher orders, however, a sudden singularity must be present, which is triggered by the (non-vanishing) second-time derivatives of the scalar-tensor potential. Furthermore, using an explicit model supporting a sudden singularity we shall show that the corresponding pressure of the matter fields must necessarily diverge, but the strong energy condition (SEC) can be satisfied. Furthermore, we will also show these results are independent of the sectional curvature $k$, due to the fact that the terms dependent on this quantity are always sub-dominant near the sudden singularity.

This paper is organized in the following manner: in Sec. \ref{sec:theory}, we introduce the generalized hybrid metric-Palatini gravity theory in both the geometrical and the scalar-tensor representation. In Sec. \ref{sec:singular}, taking into account the Friedmann-Lemaître-Robertson-Walker (FLRW) spacetime, we show that sudden singularities can only appear via derivatives of order four or higher in the scale factor. In Sec. \ref{sec:model}, we propose a solution for a cosmology with a sudden singularity in the fourth-order time derivative of the scale factor and use the experimental measurements of the Hubble and deceleration parameters to impose constraints on the sudden singularity time. Finally, in Sec. \ref{sec:concl} we summarize and discuss our results.

\section{Theory and equations of motion}\label{sec:theory}

In this section we will derive the equations of motion of both the geometrical and the scalar-tensor representation of the generalized hybrid metric-Palatini gravity using the variational method. For a more detailed calculation, we refer the reader to Sec. 2.3 and Appendix A of Ref. \cite{thesisJL}.

\subsection{Geometrical representation}

The action that describes the generalized hybrid metric-Palatini gravity is given by
\begin{equation}\label{actiongeo}
S=\frac{1}{2\kappa^2}\int_\Omega\sqrt{-g}f\left(R,\cal{R}\right)d^4x+S_m,
\end{equation}
where $\kappa^2\equiv 8\pi G/c^4$, $G$ is the Newtonian gravitational constant, $c$ is the speed of light, $\Omega$ is the spacetime volume,  and $g$ is the determinant of the spacetime metric $g_{ab}$. The Lagrangian density $f(R,\mathcal{R})$ is an arbitrary function of two scalars: $R\equiv g_{\mu\nu}R^{\mu\nu}$, which is the usual Ricci scalar of the metric $g_{ab}$ built with its Christoffel symbols, and $\mathcal{R} \equiv g_{ab}\mathcal{R}^{ab}$, which is the Palatini Ricci scalar of an independent connection $\hat\Gamma^c_{ab}$, i.e.,  its Ricci tensor $\mathcal R_{ab}$ is defined as
\begin{equation}\label{palatiniriccidef}
\mathcal{R}_{ab}=\partial_c \hat\Gamma^c_{ab}-\partial_b\hat\Gamma^c_{ac}+ \hat\Gamma^c_{cd}\hat\Gamma^d_{ab}-\hat\Gamma^c_{ad}\hat\Gamma^d_{cb},
\end{equation}
where $\partial_a$ denotes a partial derivative with respect to the coordinates $x^a$. The meaning that the connection $\hat\Gamma^c_{ab}$ is independent is that it is not necessarily metric-compatible, though we assume it to be symmetric (i.e., torsionless). Finally, $S_m$ is the matter action defined as $S_m=\int d^4x\sqrt{-g}\;{\mathcal L}_m$, where ${\mathcal L}_m$ is the matter Lagrangian density considered to be minimally coupled to the metric $g_{ab}$. In this work, we shall consider a system of units for which $G=c=1$, and thus $\kappa^2=8\pi$.

The gravitational sector of the action \eqref{actiongeo} has two independent variables, namely the metric $g_{ab}$ and the connection $\hat\Gamma^c_{ab}$, and thus we can obtain two equations of motion from a variational principle. A variation with respect to the metric $g_{ab}$ yields the modified field equations
\begin{equation}\label{genfield}
\frac{\partial f}{\partial R}R_{ab}+\frac{\partial f}{\partial\mathcal{R}}\mathcal{R}_{ab}-\frac{1}{2}g_{ab}f\left(R,\cal{R}\right)   
-\left(\nabla_a\nabla_b-g_{ab}\Box\right)\frac{\partial f}{\partial R}=\kappa^2 T_{ab},
\end{equation}
where $\nabla_a$ is the covariant derivative and $\Box=\nabla^a\nabla_a$ the d'Alembert operator, both defined in terms of the metric $g_{ab}$, and $T_{ab}$ is the stress-energy tensor defined in the usual manner as
\begin{equation}\label{defSET}
T_{ab}=-\frac{2}{\sqrt{-g}}\frac{\delta(\sqrt{-g}\,{\mathcal L}_m)}{\delta(g^{ab})}.
\end{equation}

On the other hand, a variation of  the action \eqref{actiongeo} with respect to the independent connection $\hat\Gamma^c_{ab}$ provides the relation
\begin{equation}\label{eomgamma}
\hat\nabla_c\left(\sqrt{-g}\frac{\partial f}{\partial\cal{R}}g^{ab}\right)=0 ,
\end{equation}
where $\hat\nabla_a$ is the covariant derivative written in terms of the independent connection
$\hat\Gamma^c_{ab}$, leaving the partial derivatives unchanged. As $\sqrt{-g}$ is a scalar density of weight 1, this implies that $\hat\nabla_c \sqrt{-g}=0$ (see Sec. B.1 of Ref. \cite{thesisJL} for an explicit derivation of this result), so that Eq. \eqref{eomgamma} can be written as $\hat\nabla_c\left(\frac{\partial f}{\partial
\cal{R}}g^{ab}\right)=0$. This means that defining a new metric ${\hat g}_{ab}$ as
\begin{equation}\label{hab}
{\hat g}_{ab}=g_{ab} \frac{\partial f}{\partial \mathcal{R}},
\end{equation}
the independent connection $\hat\Gamma^c_{ab}$ becomes the Levi-Civita connection for the metric ${\hat g}_{ab}$, i.e., it can be written in terms of ${\hat g}_{ab}$ in the form
\begin{equation}
\hat\Gamma^a_{bc}=\frac{1}{2}\hat g^{ad}\left(\partial_b {\hat g}_{dc}+\partial_c {\hat g}_{bd}-\partial_d {\hat g}_{bc}\right).
\end{equation}

As the metrics $ {\hat g}_{ab}$ and $g_{ab}$ are conformally related with a conformal factor ${\partial f}/{\partial \mathcal{R}}$, this implies that the two Ricci tensors $R_{ab}$ and $\mathcal R_{ab}$, which were assumed to be independent, are related to each other by
\begin{equation}\label{riccirel}
\mathcal R_{ab}=R_{ab}-\frac{1}{f_\mathcal R}\left(\nabla_a\nabla_b+\frac{1}{2}g_{ab}\Box\right)f_\mathcal
R+\frac{3}{2f_\mathcal R^2}\partial_a f_\mathcal R\partial_bf_\mathcal R,
\end{equation}
where $f_{\mathcal{R}} \mathcal  \equiv \partial f/\partial \mathcal{R}$. Equations \eqref{eomgamma} and \eqref{riccirel} are equivalent, and thus one usually uses the latter due to its simpler structure.

\subsection{Scalar-tensor representation}

Similarly to (both metric and Palatini) $f\left(R\right)$ gravity, the generalized hybrid metric-Palatini theory can be rewritten in terms of a dynamically equivalent scalar-tensor representation. This can be achieved via the introduction of two auxiliary fields $\alpha$ and $\beta$ in the action of Eq. \eqref{actiongeo} in the form \cite{Bombacigno:2019did}
\begin{equation}
S=\frac{1}{2\kappa^2}\int_\Omega \sqrt{-g}\Big[f\left(\alpha,\beta\right)+\frac{\partial f}{\partial \alpha}\left(R-\alpha\right)   
+\frac{\partial f}{\partial\beta}\left(\cal{R}-\beta\right)\Big]d^4x+S_m.\label{gensca}
\end{equation}
One can easily see from action \eqref{gensca} that the particular solution $\alpha=R$ and $\beta=\mathcal R$ reduces to the original action of the geometrical representation, given in Eq. \eqref{actiongeo}. Defining two scalar fields $\varphi$ and $\psi$ as the partial derivatives of the function $f\left(R,\mathcal R\right)$ with respect to $R$ and $\mathcal R$ respectively, i.e., $\varphi=\partial f/\partial\alpha$ and $\psi=-\partial f/\partial\beta$, where the negative sign in $\psi$ is imposed to guarantee the regularity of its kinetic term, Eq. \eqref{gensca} becomes
\begin{equation} \label{action3}
S=\frac{1}{2\kappa^2}\int_\Omega \sqrt{-g}\left[\varphi R-\psi\mathcal{R}-V\left(\varphi,\psi\right)\right]d^4x,
\end{equation}
where the function $V\left(\varphi,\psi\right)$ plays the role of the interaction potential of the scalar fields $\varphi$ and $\psi$. This function is defined in terms of the Lagrangian density $f\left(R,\mathcal R\right)$ and the auxiliary fields $\alpha$ and $\beta$ as
\begin{equation}\label{potential}
V\left(\varphi,\psi\right)=-f\left(\alpha,\beta\right)+
\varphi\alpha-\psi\beta.
\end{equation}

Taking into account the relation between $R_{ab}$ and $\mathcal R_{ab}$ deduced in Eq. \eqref{riccirel}, one can find a relationship between $R$ and $\mathcal R$ by taking a trace with $g^{ab}$. In addition to this, since the conformal factor can be written as $f_\mathcal R=-\psi$, we obtain
\begin{equation}\label{confrt}
\mathcal{R}=R-\frac{3}{\psi}\Box\psi+\frac{3}{2\psi^2}\partial^a \psi\partial_a \psi.
\end{equation}
Finally, inserting Eq. \eqref{confrt} into the action \eqref{action3} to cancel the dependency in $\mathcal R$ and discarding the term $\Box\psi$ as it does not contribute to the field equations (as it can always be turned into a boundary term via the Stokes theorem, which vanishes by definition as the variation is zero at the boundary), we obtain the action for the scalar-tensor representation of the generalized hybrid metric-Palatini gravity as
\begin{equation}
S=\frac{1}{2\kappa^2}\int_\Omega \sqrt{-g}\Big[\left(\varphi-\psi\right) R-\frac{3}{2\psi}\partial^a\psi\partial_a\psi    
-V\left(\varphi,\psi\right)\big]d^4x+S_m.\label{actionsca}
\end{equation}

The action in Eq. \eqref{actionsca} is now a function of three independent variables, namely the metric $g_{ab}$ and the scalar fields $\varphi$ and $\psi$. Thus, we can obtain three equations of motion. Variation of Eq. \eqref{actionsca} with respect to the metric $g_{ab}$ provides the modified field equations
\begin{eqnarray}
\left(\varphi-\psi\right) G_{ab}&=&\kappa^2T_{ab}+\frac{3}{2\psi}\partial_a\psi\partial_b\psi-\frac{3}{4\psi}g_{ab}\partial^c\psi\partial_c\psi	\nonumber  \\
&& -\frac{1}{2}Vg_{ab}+\left(\nabla_a\nabla_b-g_{ab}\Box\right)\left(\varphi-\psi\right)\,.\label{fieldsca}
\end{eqnarray}

Varying the action in Eq.\eqref{actionsca} with respect to the scalar fields $\varphi$ and $\psi$ and manipulating the resultant equations to isolate the terms $\Box\varphi$ and $\Box\psi$, yields the Klein-Gordon-like equations for these scalar fields as
\begin{eqnarray}
&&\Box\varphi+\frac{1}{3}\left(2V-\psi V_\psi-\varphi V_\varphi\right)=\frac{\kappa^2T}{3}\label{kgphi}, \\
&&\Box\psi-\frac{1}{2\psi}\partial^a\psi\partial_a\psi-
\frac{\psi}{3}\left(V_\varphi+V_\psi\right)=0 \label{kgpsi},
\end{eqnarray}
where $T=g^{ab}T_{ab}$ is the trace of the stress-energy tensor. A final equation that will be useful is the relationship between the derivative of the potential with respect to the field $\varphi$ and the Ricci scalar $R$. This equation is obtained from the direct variation of Eq. \eqref{actionsca} with respect to the scalar field $\varphi$, which yields
\begin{equation}\label{potrel}
R=V_\varphi.
\end{equation}

In the following sections, we work in this scalar-tensor representation of the theory. The system of equations of motion is thus given by Eqs. \eqref{fieldsca}--\eqref{kgpsi}.

\section{Sudden singularities in Friedman-Lemaître-Robertson-Walker (FLRW) cosmologies}\label{sec:singular}

\subsection{Equations of motion in a FLRW background}

In this work, as mentioned in the Introduction, we are interested in studying the appearance of sudden singularities in a cosmological context. Here, it is usual to assume that the spacetime is homogeneous and isotropic. Spacetimes with these properties are described by the FLRW line element, given in spherical coordinates $\left(t,r,\theta,\phi\right)$ by
\begin{equation}\label{flrw}
ds^2=-dt^2+a^2\left(t\right)\left(\frac{dr^2}{1-kr^2}+r^2d\Omega^2\right),
\end{equation}
where $t$ is the proper time coordinate, $r$ is the radial coordinate, $a\left(t\right)$ is the scale factor, assumed to be a function of time only to preserve the homogeneity of the spacetime, $k$ is the sectional curvature and takes the values $\{0,\pm 1\}$ depending on the spacetime geometry of the constant time slices, and $d\Omega^2=r^2d\theta^2+r^2\sin^2\theta d\phi^2$ is the line element over the unit sphere surface.

We shall also consider that the matter distribution is described by an isotropic perfect fluid, i.e., the stress-energy tensor $T_{ab}$ can be written in the diagonal form
\begin{equation}\label{stress}
g^{ab}T_{ab}=\text{diag}\left(-\rho,p,p,p\right),
\end{equation}
where $\rho$ is the energy density and $p$ is the pressure, both assumed to be only functions of time $t$. The conservation of energy for this distribution of matter is given by the usual statement $\nabla_aT^{ab}=0$, computed with the covariant derivative of the metric.

Introducing Eqs. \eqref{flrw} and \eqref{stress} into the modified field equations in Eq. \eqref{fieldsca} and assuming that the scalar fields $\varphi$ and $\psi$ are solely functions of $t$, one verifies that there are only two independent equations of motion, arising from the $\{t,t\}$ and $\{r,r\}$ components. These equations are
\begin{equation}\label{field1}
3\frac{\dot a^2+k}{a^2}\left(\varphi-\psi\right)+3\frac{\dot a}{a}\left(\dot\varphi-\dot\psi\right)-\frac{3\dot\psi^2}{4\psi}-\frac{V}{2}=8\pi\rho,
\end{equation}
\begin{equation}
\left(2\frac{\ddot a}{a}+\frac{\dot a^2}{a^2}+\frac{k}{a^2}\right)\left(\varphi-\psi\right)+2\frac{\dot a}{a}\left(\dot\varphi-\dot\psi\right)
+\frac{3\dot\psi^2}{4\psi}+\ddot\varphi-\ddot\psi-\frac{V}{2}=-8\pi p \,,
\label{field2}
\end{equation}
respectively.
On the other hand, the equations of motion for the scalar fields in the background of Eq. \eqref{flrw} become
\begin{eqnarray}
\ddot\varphi+3\frac{\dot a}{a}\dot\varphi-\frac{1}{3}\left(2V-\varphi V_\varphi-\psi V_\psi\right)&=&\frac{8\pi}{3}\left(\rho-3p\right), \label{eqphi} \\
\ddot\psi+3\frac{\dot a}{a}\dot\psi-\frac{\dot\psi^2}{2\psi}+\frac{\psi}{3}\left(V_\varphi+V_\psi\right)&=&0. \label{eqpsi}
\end{eqnarray}
Finally, the conservation equation for the stress-energy tensor discussed above, $\nabla_aT^{ab}=0$, obviously yields the same equation for the perfect fluid as in GR, that is
\begin{equation}\label{eqmat}
\dot\rho=-3\frac{\dot a}{a}\left(\rho+p\right).
\end{equation}
Note that Eq. \eqref{field1} is not independent of the remaining equations. To prove that, one verifies that taking a derivative of Eq. \eqref{field1} with respect to time and using Eqs. \eqref{eqphi}, \eqref{eqpsi}, \eqref{eqmat}, and \eqref{field2} to cancel the terms depending on $\ddot\varphi$, $\ddot\psi$, $\dot\rho$, $\ddot a$, respectively, one recovers Eq. \eqref{field1}. Thus, in the following sections, we shall explore the possibility of the appearance of sudden singularities in the system of Eqs. \eqref{field2}--\eqref{eqmat}.

Finally, let us also write explicitly the form of Eq. \eqref{potrel} in the background of Eq. \eqref{flrw}, which is
\begin{equation}\label{potflrw}
V_\varphi=6\left(\frac{\ddot a}{a}+\frac{\dot a^2}{a^2}+\frac{k}{a^2}\right).
\end{equation}
Again, note that Eq. \eqref{potflrw} is not independent of the system of Eqs. \eqref{field2}--\eqref{eqmat}, and thus it shall only be used as a tool to simplify the upcoming equations by writing dependencies on $V_\varphi$ in terms of the scale factor $a$ and its time derivatives.

\subsection{Sudden singularities in the scale factor}\label{sec:sing1}

A sudden singularity is defined as a future event in the cosmological evolution in which the scale factor, the expansion rate, and the energy density are all finite, but the pressure diverges, thus inducing a divergence in higher-order derivatives of the scale factor \cite{Barrow:2004xh}. Let us consider the possibility of a sudden singularity arising in the system of Eqs. \eqref{field2}--\eqref{eqmat}. In this scenario, the scale factor $a$ and its first time derivative $\dot a$ remain finite, but higher-order time derivatives like $\ddot a$ are allowed to diverge in a finite time. For Eq. \eqref{field2} to be satisfied throughout this time evolution, one verifies that a divergence in $\ddot a$ must be compensated by a symmetric divergence of the same degree in some other variable\footnote{In this analysis it is implicitly assumed that the target functions are smooth enough, i.e., at least $C^4$-differentiable in time.}. In order to have a better insight on which variables should be allowed to compensate the divergence in $\ddot a$, let us rewrite Eqs. \eqref{field1} and \eqref{field2} in terms of an effective energy density $\rho_{\text{eff}}$ and an effective pressure $p_{\text{eff}}$ in the form
\begin{equation}
3\frac{\dot a^2+k}{a^2}=8\pi\rho_{\text{eff}},
\end{equation}
\begin{equation}
2\frac{\ddot a}{a}+\frac{\dot a^2}{a^2}+\frac{k}{a^2}=-8\pi p_{\text{eff}},
\end{equation}
where $\rho_\text{eff}$ and $p_\text{eff}$ are defined as
\begin{equation}
\rho_\text{eff}=\frac{1}{\varphi-\psi}\left\{\rho-\frac{1}{8\pi}\left[3\frac{\dot a}{a}\left(\dot\varphi-\dot\psi\right)-\frac{3\dot\psi^2}{4\psi}-\frac{V}{2}\right]\right\},
\end{equation}
\begin{equation}
p_\text{eff}=\frac{1}{\varphi-\psi}\left\{p-\frac{1}{8\pi}\left[2\frac{\dot a}{a}\left(\dot\varphi-\dot\psi\right)+\frac{3\dot\psi^2}{4\psi}+\ddot\varphi-\ddot\psi+\frac{V}{2}\right]\right\}.
\end{equation}
We thus verify that the effective energy density is proportional to the scalar fields and their first-order time derivatives, whereas the effective pressure features terms proportional to the second-order derivatives of the scalar fields. According to the definition of sudden singularity, to keep both $\rho$ and $\rho_\text{eff}$ finite and allow for $p$ and $p_\text{eff}$ to diverge, the quantities $\varphi$, $\psi$, $\dot\varphi$ and $\dot\psi$ should remain finite, which consequently implies that $V$ is finite, and the quantities $\ddot\varphi$ and $\ddot\psi$ should be allowed to diverge. Thus, one concludes that the divergence in $\ddot a$ must be compensated by divergences in $\ddot\varphi$, $\ddot\psi$, or $p$.

Turning now to the equations for the scalar fields, from Eq. \eqref{eqphi} one verifies that, keeping the same assumptions as in the previous paragraph, a divergence in $\ddot\varphi$ must be compensated by a divergence in the pressure $p$. On the other hand, from Eq. \eqref{eqpsi} one verifies that the only possibly divergent term in this equation is $\ddot\psi$. As this equation cannot be satisfied if none of the remaining terms compensates for the divergence of $\ddot\psi$, one concludes that although we have not excluded the possibility of a divergence in $\ddot\psi$, the structure of Eq. \eqref{eqpsi} prevents the occurrence of such a divergence. Consequently, as the divergence in $\ddot a$ from Eq. \eqref{field2} cannot be compensated by a divergence in $\ddot\psi$, the only possibility left is that both $\ddot\varphi$ and $p$ must also present divergences (sudden singularities) at the same instant as $\ddot a$.

Finally, let us look into the energy conservation equation. From Eq. \eqref{eqmat}, one verifies that a divergence in the pressure $p$ must be compensated either by a divergence in the energy density $\rho$ or its first time derivative $\dot\rho$. As we want to preserve the regularity of the energy density, i.e., that $\rho$ remains finite throughout the entire time evolution, we conclude that $\dot\rho$ must also present a sudden singularity at the same instant and with the same degree of divergence as $\ddot a$.

\subsubsection{Absence of sudden singularities in $\ddot a$}\label{sec:nosing1}

Given the arguments stated above, we shall be looking for sudden singularities occurring in $\ddot a$, $\ddot\varphi$, $\dot\rho$ and $p$, while keeping $a$, $\dot a$, $\rho$, $\varphi$, $\dot\varphi$, $\psi$, $\dot\psi$ and $\ddot\psi$ finite. Let us assume that these finite-time divergences occur at an instant $t=t_s$. Thus, as $t\to t_s$, the terms depending on non-divergent quantities are sub-dominant and one can derive an asymptotic system of equations close to $t=t_s$.  We are thus left with the asymptotic system of Eqs. \eqref{field2}, \eqref{eqphi} and \eqref{eqmat}, as
\begin{equation}
\frac{\ddot a}{a}\simeq -\frac{8\pi p + \ddot \varphi}{2\left(\varphi-\psi\right)},\label{aseq1}
\end{equation}
\begin{equation}
\ddot\varphi\simeq -8\pi p,\label{aseq2}
\end{equation}
\begin{equation}
\dot\rho\simeq 3p\frac{\dot a }{a}.\label{aseq3}
\end{equation}

Now, the dependency of Eq. \eqref{aseq1} in the pressure $p$ can be cancelled via the use of Eq. \eqref{aseq2}. Consequently, one obtains an identically vanishing right-hand side in Eq. \eqref{aseq1}. Similarly to what happens with Eq. \eqref{eqpsi}, one verifies that there are no other divergent terms in this equation that could compensate a divergence in $\ddot a$, and thus a sudden singularity in $\ddot a$ is prevented by the very structure of the field equations in the generalized hybrid metric-Palatini gravity. This result is somewhat expected from the analysis in Ref. \cite{Barrow:2019cuv}, as the scalar field $\varphi$ is dynamically equivalent to a Brans-Dicke scalar field with a parameter $\omega_{BD}=0$, a theory in which one would obtain $\ddot a=0$.

\subsubsection{Absence of sudden singularities in $\dddot a$}\label{sec:nosing2}

The results of Sec. \ref{sec:nosing1} indicate that the terms $\ddot a$ and $\ddot\psi$, which were allowed to diverge in a finite time $t_s$, must remain finite to preserve the consistency of Eqs. \eqref{field2} and \eqref{eqpsi}, respectively. However, these results do not yet exclude the possibility of sudden singularities in $\ddot\phi$, $\dot\rho$, $p$, or even higher-order time derivatives of the scale factor, like $\dddot a$. To explore this possibility, one can differentiate the system of Eqs. \eqref{field2}--\eqref{eqmat} with respect to time and repeat the analysis.

Let us start by analyzing the first time derivative of Eq. \eqref{eqpsi}. Similarly as before, as we have already concluded that the term $\ddot\psi$ must be finite, the only term that could possibly diverge in this equation is $\dddot\psi$. In the absence of a second divergent term to compensate for the divergence in $\dddot\psi$ and preserve the validity of this equation, we are forced to conclude that $\dddot\psi$ must remain finite. The remaining equations, namely, Eqs. \eqref{field2}, \eqref{eqphi} and \eqref{eqmat} yield, after differentiation with respect to time, the following asymptotic system near $t\to t_s$
\begin{equation}\label{daseq1}
\dddot a\simeq -\frac{16\pi p\dot a+4\dot a\ddot\varphi+a\left(8\pi\dot p+\dddot\varphi\right)}{2\left(\varphi-\psi\right)},
\end{equation}
\begin{equation}\label{daseq2}
\dddot\varphi\simeq\frac{8\pi}{3}\left(\dot\rho-3\dot p\right)-3\frac{\dot a}{a} \ddot\varphi,
\end{equation}
\begin{equation}\label{daseq3}
\ddot\rho\simeq -\frac{3}{a^2}\left[a\dot a\left(\dot p + \dot \rho\right)+p\left(a\ddot a-\dot a^2\right)\right].
\end{equation}

One can now use Eq. \eqref{aseq3} to cancel the dependency of Eq. \eqref{daseq2} in $\dot\rho$, following by the use of the result to cancel the dependency of Eq. \eqref{daseq1} in $\dot p$, as well as use Eq. \eqref{aseq2} to cancel the dependency in $p$. As a result, we obtain an identically vanishing right-hand side in Eq. \eqref{daseq1}. Again, as there are no other divergent terms in this equation to compensate for a possible divergence in $\dddot a$ and preserve the validity of the equation, we conclude that the structure of the field equations prevents the appearance of sudden singularities in $\dddot a$.

\subsubsection{Sudden singularities in higher-order time derivatives}

The first non-trivial results arise when one considers the fourth-order time derivatives of the scale factor, which we will be representing by $a^{(4)}$ to lighten the dot notation. To understand the origin of this non-triviality, let us consider the time derivatives of the terms depending on the potential or its partial derivatives with respect to the scalar fields $\varphi$ or $\psi$. The first and second time derivatives of the potential terms are, respectively:
\begin{equation}
\dot V=V_\varphi\dot\varphi+V_\psi\dot\psi,
\end{equation}
\begin{equation}
\ddot V=V_{\varphi\varphi}\dot\varphi^2+V_\varphi\ddot\varphi+V_{\psi\psi}\dot\psi^2+V_\psi\ddot\psi+2V_{\varphi\psi}\dot\varphi\dot\psi.
\end{equation}
Thus, although the potential $V$ itself is finite throughout the entire time evolution, along with the scalar fields $\varphi$ and $\psi$ and their time derivatives $\dot\varphi$, $\dot\psi$ and $\ddot\psi$, the second time derivatives of the potential-dependent terms contribute with a possibly divergent term $V_\varphi\ddot\varphi$. The appearance of these extra terms when one takes the second time derivative of the system of Eqs. \eqref{field2}-\eqref{eqpsi} effectively breaks the structure of the field equations that prevent the sudden singularities from appearing in the time derivatives of the scale factor, i.e., the self-consistency of the field equations that leads to a cancellation of the divergent terms in the higher-order derivatives of $a$ by the same-order derivatives of $\varphi$ (e.g. when Eqs.\eqref{aseq2} and \eqref{daseq2} were introduced into Eqs.\eqref{aseq1} and \eqref{daseq1}, respectively, leading to trivial identities) no longer holds when the extra terms arise.

Let us then analyze the second time derivatives of the system of Eqs. \eqref{field2}--\eqref{eqmat}. In particular, note that the second time derivative of Eq. \eqref{eqpsi} will now feature a term proportional to $\ddot\varphi$ which can compensate a possible divergence in the fourth time derivative $\psi^{(4)}$, and this equation will no longer be trivial in the asymptotic $t\to t_s$ regime. Taking the second time derivatives of Eqs. \eqref{field2}--\eqref{eqmat} and discarding sub-dominant terms near $t_s$ (i.e., those that have already been proven to remain finite), one obtains
\begin{eqnarray}
a^{(4)} & \simeq & -\frac{1}{4\left(\varphi-\psi\right)}\left[64\pi p\left(\dot a^2+a \ddot a\right)+64\pi a\dot a\dot p+16\pi a^2\ddot p+\ddot\varphi\left(2k+14\dot a^2+16a\ddot a+a^2V_\varphi\right)\right.\nonumber \\
&& \qquad \qquad \left.+12a\dot a\dddot\varphi+2a^2\left(\varphi^{(4)}-\psi^{(4)}\right)\right],\label{ddaseq1}
\end{eqnarray}
\begin{equation}
\varphi^{(4)}  \simeq \frac{8\pi}{3}\left(\ddot\rho-3\ddot p\right)-3\frac{\dot a}{a}\dddot\varphi
\quad+\ddot\varphi\left[6\left(\frac{\dot a^2}{a^2}-\frac{\ddot a}{a}\right)+\frac{1}{3}\left(V_\varphi-\varphi V_{\varphi\varphi}-\psi V_{\varphi\psi}\right)\right],\label{ddaseq2}
\end{equation}
\begin{equation}\label{ddaseq3}
\psi^{(4)}\simeq-\frac{\psi\ddot\varphi}{3}\left(V_{\varphi\varphi}+V_{\psi\varphi}\right),
\end{equation}
\begin{equation}
\rho^{(3)}  \simeq  -3\frac{\dot a}{a}\left(\ddot p+\ddot \rho\right)+6\left(\frac{\dot a^2}{a^2}-\frac{\ddot a}{a}\right)\left(\dot p + \dot \rho\right)
	-6p\left(\frac{\dot a^3}{a^3}-\frac{3\dot a\ddot a}{2a^2}+\frac{\dddot a}{2a}\right).\label{ddaseq4}
\end{equation}

One can now cancel the dependencies of Eq. \eqref{ddaseq1} in the variables $p$, $\dot p$, $\ddot p$, $\psi^{(4)}$ and $V_\varphi$ using Eqs. \eqref{aseq2}, \eqref{daseq2}, \eqref{ddaseq2}, \eqref{ddaseq3} and \eqref{potflrw}, respectively. As a result, we obtain the consistency equation
\begin{equation}\label{suddena4}
\frac{6a^{(4)}}{a}\simeq\ddot\varphi\ V_{\varphi\varphi}.
\end{equation}

This result confirms that the sudden singularity is sourced by the extra terms arising from the second time derivatives of the potential-dependent terms, as taking the particular case of a vanishing potential would lead again to an identically vanishing right-hand side of Eq. \eqref{ddaseq1}, and consequently $a^{(4)}$ would be forcefully finite. Furthermore, notice that the potential $V$ must be at least quadratic in the scalar field $\varphi$ to guarantee that the right-hand side of Eq. \eqref{suddena4} is non-zero.

\section{Model with sudden singularities}\label{sec:model}

\subsection{Explicit solution for the scale factor}\label{sec:solution}

Considering the results of Eq. \eqref{suddena4}, we now analyze models with the simplest possible $V$ that supports the existence of sudden singularities while simplifying Eqs. \eqref{field2}--\eqref{eqmat}. This potential is quadratic in both $\varphi$ and $\psi$, of the form
\begin{equation}\label{potential1}
V\left(\varphi,\psi\right)=V_0\left(\varphi-\psi\right)^2,
\end{equation}
where $V_0$ is an arbitrary constant. As shown in previous works \cite{Rosa:2017jld,Rosa:2018jwp}, the potential chosen in Eq. \eqref{potential1} is particularly useful given the fact that all potential-dependent terms in Eqs. \eqref{eqphi} and \eqref{eqpsi} cancel mutually, and these two equations simplify considerably. Consequently, the consistency equation  \eqref{suddena4} becomes
\begin{equation}\label{cons1}
\frac{a^{(4)}}{a}=\frac{1}{3}V_0\ddot\varphi.
\end{equation}
Note that this result requires the sudden singularities to occur simultaneously in both $a^{(4)}$ and $\ddot\varphi$. As the system of Eqs. \eqref{field2}--\eqref{eqmat} consists of 4 independent equations for the 6 unknowns $a$, $\varphi$, $V$, $\psi$, $\rho$, and $p$, and we have already fixed the form of the potential $V$ in Eq. \eqref{potential1}, we still have the freedom to impose a constraint on the system, such as an explicit form for the scale factor, to close the system. For the sake of this paper, and following Barrow \cite{Barrow:2004xh}, let us choose the following ansatz for the scale factor
\begin{equation}\label{scale1}
a\left(t\right)=\left(a_s-1\right)\left(\frac{t}{t_s}\right)^\gamma+1-\left(1-\frac{t}{t_s}\right)^\delta,
\end{equation}
where $a_s$ is the value of the scale factor at the sudden singularity time $t_s$, and $\gamma$ and $\delta$ are constant exponents that shall be required to satisfy a few conditions, as explained next. The chosen ansatz for $a$ satisfies the property $a\left(0\right)=0$. The $n$-th derivative of Eq. \eqref{scale1} is given in the general form
\begin{equation}
a^{(n)}\left(t\right)=\left(a_s-1\right)\left(\frac{t}{t_s}\right)^{\gamma-n}t_s^{-n}\prod_{i=0}^{n-1}\left(\gamma-i\right)
 +\left(-1\right)^{n+1}\left(1-\frac{t}{t_s}\right)^{\delta-n}t_s^{-n}\prod_{i=0}^{n-1}\left(\delta-i\right).
\label{dscale1}
\end{equation}
The first term in Eq. \eqref{dscale1} is responsible for keeping the regularity of the scale factor at the origin, and thus all derivatives of this term up to order $n$ should be regular at $t=0$ if $\gamma>n$. On the other hand, the second term in Eq. \eqref{dscale1} causes the sudden singularity to appear at $t=t_s$. These singularities will appear whenever $n>\delta$. Furthermore, the exponents $\gamma$ and $\delta$ must not be whole numbers to avoid the situations $\gamma=n$ or $\delta=n$ to arise, as this would remove the dependencies of $a^{(n)}$ in either $t/t_s$ or $1-t/t_s$, respectively. As we want the scale factor in Eq. \eqref{scale1} to present a sudden singularity only at the fourth-order time derivative $a^{(4)}$, we thus require that $3<\delta<4<\gamma$.

Under the considerations above, we have that $a^{(4)}\left(t\to t_s\right)\to-\infty$. According to Eq. \eqref{cons1}, $\ddot \varphi$ will diverge either to $+\infty$ or $-\infty$ depending on the sign of $V_0$ being negative or positive, respectively. Furthermore, from Eq. \eqref{aseq2} it is clear that the pressure $p$ will diverge in the opposite direction as $\ddot\varphi$. Consequently, in order to preserve the validity of the SEC, we require that $p\left(t\to t_s\right)\to +\infty$, which requires $\ddot\varphi\left(t\to t_s\right)\to-\infty$, and thus the constant $V_0$ must be positive.

\begin{figure}[t!]
\includegraphics[scale=1.1]{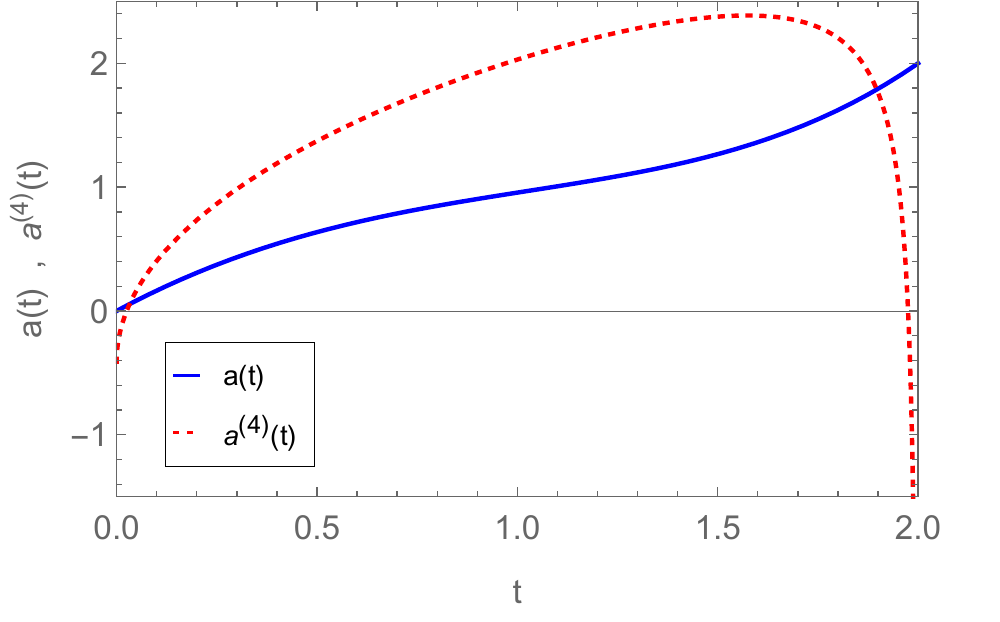}
\caption{Scale factor $a\left(t\right)$ from Eq.\eqref{scale1} and its fourth time derivative $a^{(4)}\left(t\right)$ as a function of time $t$ with $a_s=2$, $t_s=2$, $\gamma=9/2$, $\delta=7/2$. A sudden singularity in $a^{(4)}$ occurs at time $t=t_s$.}
\label{fig:scale}
\end{figure}

In Fig. \ref{fig:scale} we plot the scale factor $a\left(t\right)$ given in Eq. \eqref{scale1} and its fourth time derivative $a^{(4)}\left(t\right)$ as functions of time $t$ for a given combination of parameters. The scale factor presents an initial deceleration period which is followed by a late-time cosmic acceleration. The fourth-order time derivative starts at a negative finite value at $t=0$, changes sign and stays positive throughout most of the time evolution, until it suddenly reverses its behavior and diverges to negative values at $t_s$.

\subsection{Constraints from the cosmological parameters}

\begin{figure*}[t!]
\includegraphics[scale=0.7]{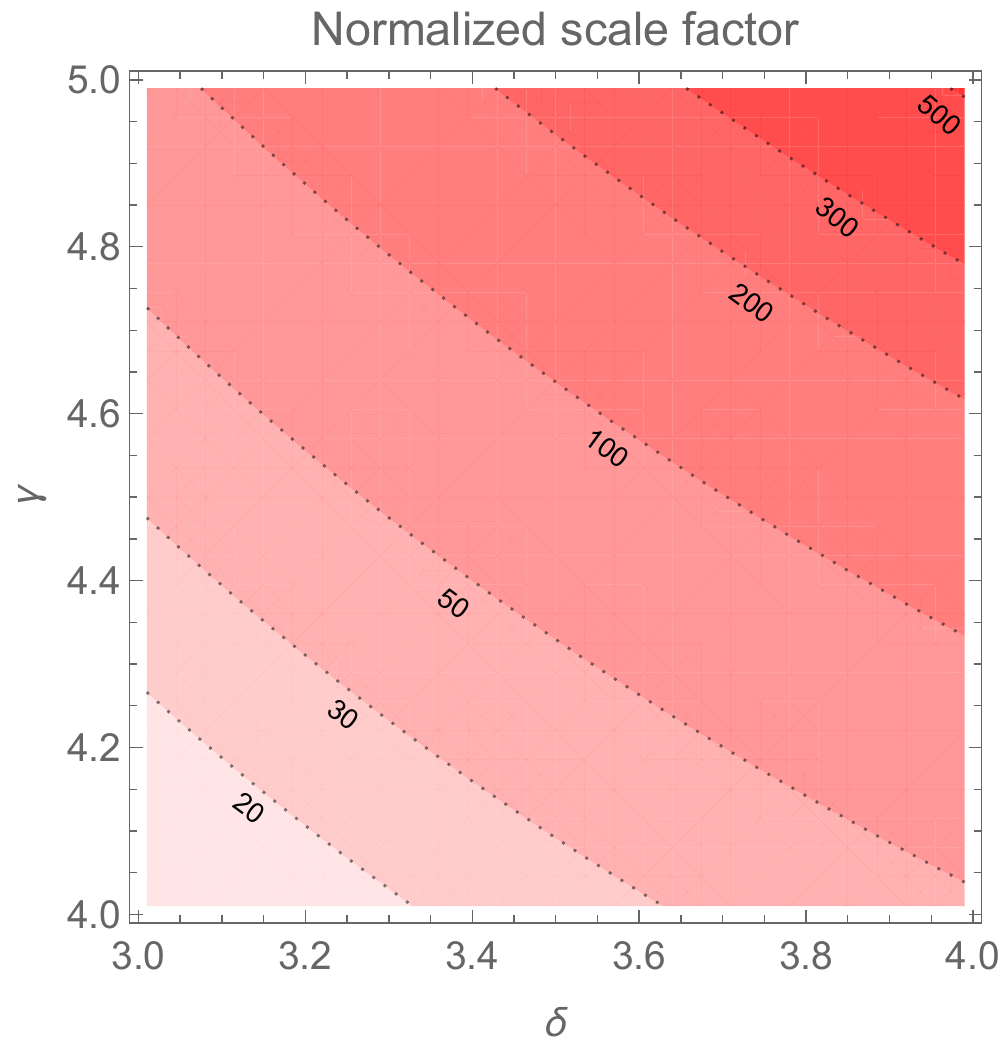}
\ \ \ \ \ \ \ \ \ \
\includegraphics[scale=0.7]{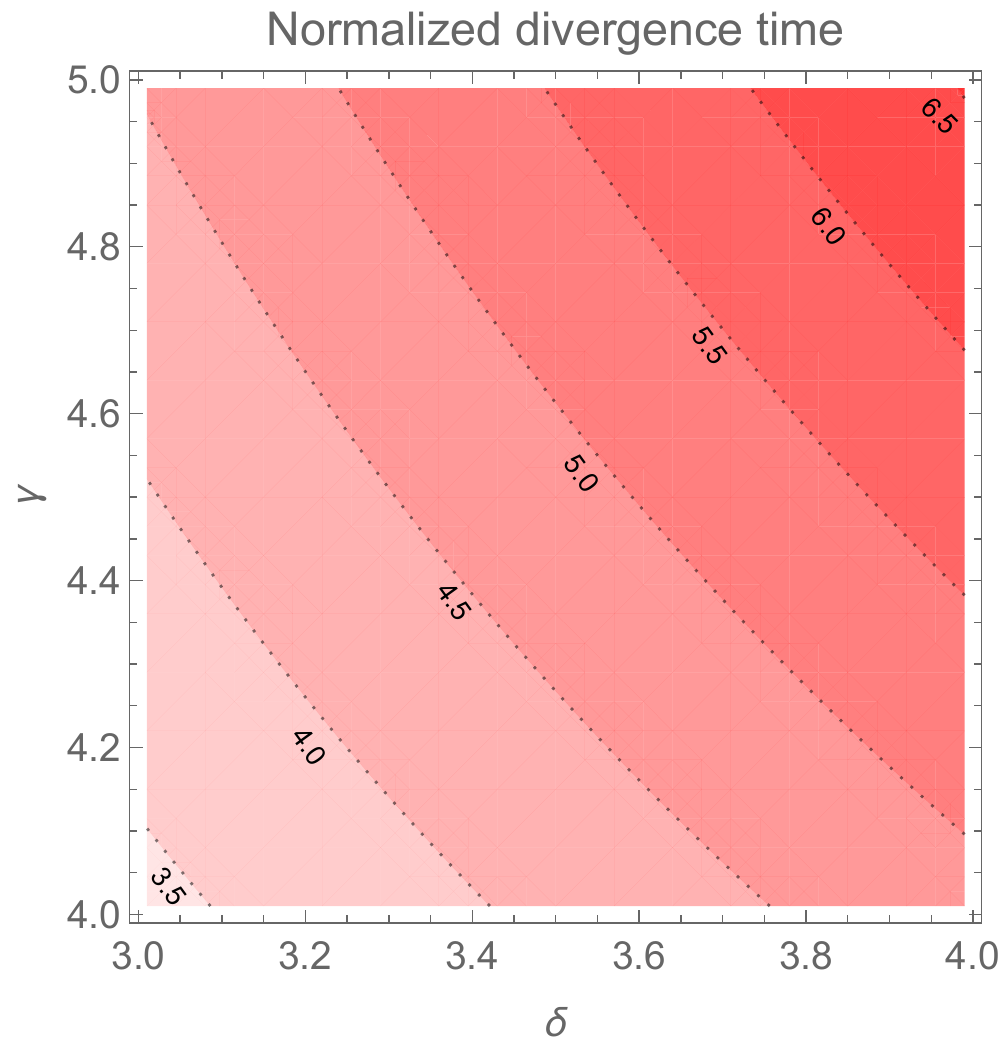}
\caption{Normalized scale factor $\bar a=a_s/a_0$ (left panel) and normalized divergence time $\bar t=t_s/t_0$ (right panel) as a function of $\gamma$ and $\delta$ for the model given in Eq.\eqref{scale1} subjected to the cosmological parameter constraints in Eq.\eqref{cosmotest}. Both $\bar a$ and $\bar t$ are shown to grow with $\gamma$ and $\delta$.}
\label{fig:cosmos}
\end{figure*}

In this section, we use the measurements of the cosmological parameters from the Planck Satellite \cite{Aghanim:2018eyx} to impose constraints on the value of the scale factor at the sudden singularity event, $a_s$, and the time $t_s$ at which it occurs. In particular, we will be working with the Hubble parameter $H$ and the deceleration parameter $q$ defined in terms of the scale factor and its time derivatives as
\begin{equation}\label{cosmopar}
H=\frac{\dot a}{a}, \qquad\qquad q=-\frac{\ddot a a}{\dot a^2}.
\end{equation}
According to the measurements of the Planck Satellite, the experimental values of these cosmological parameters at the present time are roughly $H_0\sim 67.66\ \text{km s}^{-1} \text{Mpc}^{-1}$ and $q_0\sim -0.53$, where the value of the present time $t_0$, also known as the age of the universe, is measured to be $t_0\sim13.79\ \text{Gy}$. Using Eqs. \eqref{scale1} and \eqref{dscale1}, the cosmological parameters $H$ and $q$ from Eq. \eqref{cosmopar} take the forms
\begin{equation}\label{hubble}
H\left(t\right)=-\frac{\left(a_s-1\right)\left(t-t_s\right)\left(\frac{t}{t_s}\right)^\gamma-\delta t\left(1-\frac{t}{t_s}\right)^\delta}{t\left(t-t_s\right)\left[\left(1-\frac{t}{t_s}\right)^\delta-1-\left(a_s-1\right)\left(\frac{t}{t_s}\right)^\gamma\right]},
\end{equation}
\begin{eqnarray}\label{deceleration}
q\left(t\right)&=&-\frac{\left(1-\frac{t}{t_s}\right)^\delta-1-\left(a_s-1\right)\left(\frac{t}{t_s}\right)^\gamma}{\left[\left(a_s-1\right)\left(t-t_s\right)\left(\frac{t}{t_s}\right)^\gamma-\delta t\left(1-\frac{t}{t_s}\right)^\delta\right]^2} \times
	\nonumber \\
&&\times \left[\delta\left(\delta-1\right)t^2\left(1-\frac{t}{t_s}\right)^\delta-\gamma\left(\gamma-1\right)
\left(a_s-1\right)\left(t-t_s\right)^2\left(\frac{t}{t_s}\right)\right].
\end{eqnarray}

Inserting the measured values of the Hubble parameter $H_0$, the deceleration parameter $q_0$ and the age of the universe $t_0$ into Eqs. \eqref{hubble} and \eqref{deceleration}, i.e., writing
\begin{equation}\label{cosmotest}
H\left(t=t_0\right)=H_0,\qquad\qquad q\left(t=t_0\right)=q_0,
\end{equation}
one obtains a set of two equations for the four unknowns $\gamma$, $\delta$, $a_s$, and $t_s$. Given that in Sec. \ref{sec:solution} we have concluded that the parameters $\gamma$ and $\delta$ must satisfy the inequalities $3<\delta<4<\gamma$ to guarantee that the sudden singularity appears at the fourth-order time derivative of the scale factor $a^{(4)}$, we can set the values for $\gamma$ and $\delta$ in Eqs. \eqref{cosmotest} and solve for the two remaining unknowns $a_s$ and $t_s$. This allows us to predict what would be the divergence time $t_s$ for these models, while keeping the cosmological parameters consistent with the experimental measurements.

In Fig. \ref{fig:cosmos} we provide contour plots of the normalized scale factor $\bar a=a_s/a_0$, where $a_0=a\left(t=t_0\right)$, and normalized divergence times $\bar t=t_s/t_0$ as functions of the exponents $\gamma$ and $\delta$ in the range $3<\delta<4<\gamma<5$ under the constraint of Eq. \eqref{cosmotest}. Both $\bar a$ and $\bar t$ are shown to increase with the exponents $\gamma$ and $\delta$ in these intervals, and thus the measured values for the cosmological parameters $H_0$ and $q_0$ constraint the divergence times to the interval $t_{\rm min}<t_s<t_{\rm max}$, with $t_{\rm min}\sim 3.36t_0$ and $t_{\rm max}\sim 6.56 t_0$. The scale factor $a_s$ at which these sudden singularities happen is thus also constrained to the interval $a_{\rm min}<a_s<a_{\rm max}$, with $a_{\rm min}\sim 12.1 a_0$ and $a_{\rm max}\sim 535a_0$.

\begin{figure*}[t!]
\includegraphics[scale=0.7]{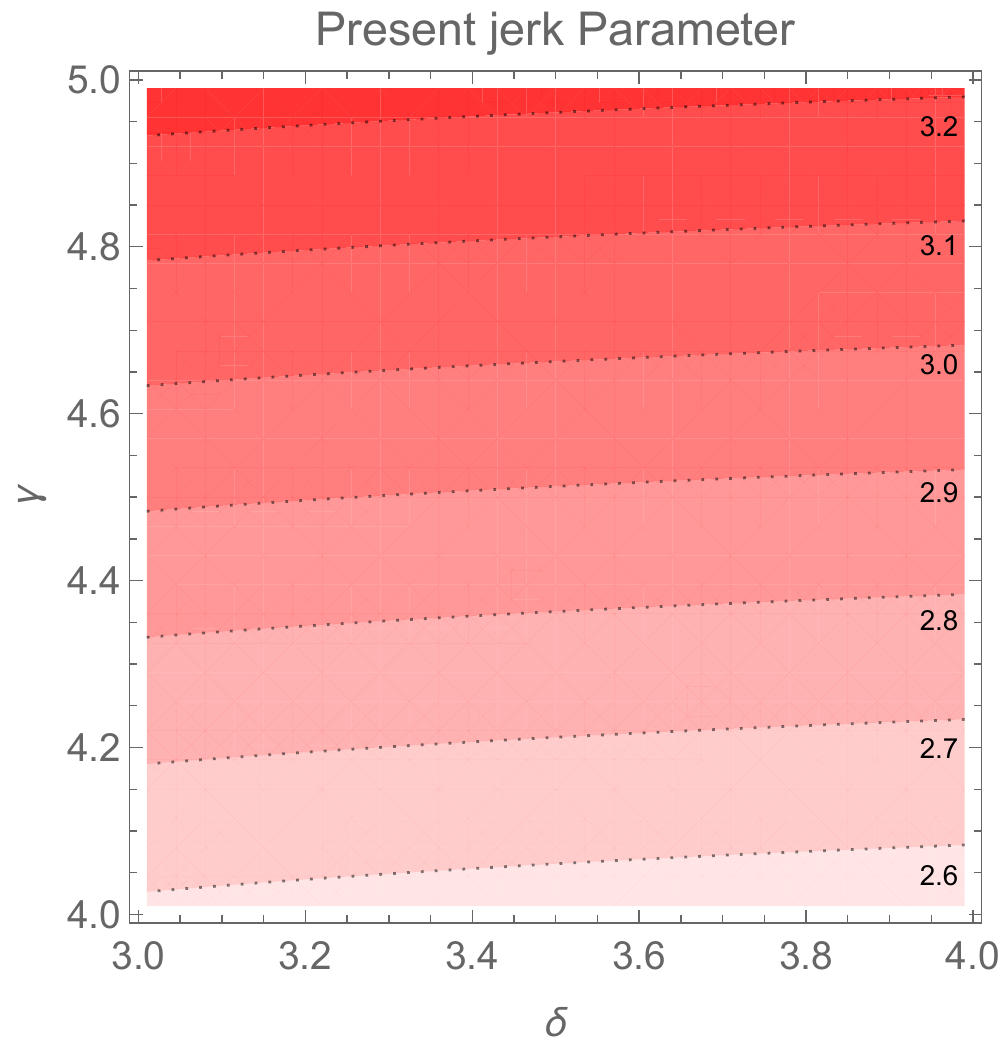}1
\includegraphics[scale=0.7]{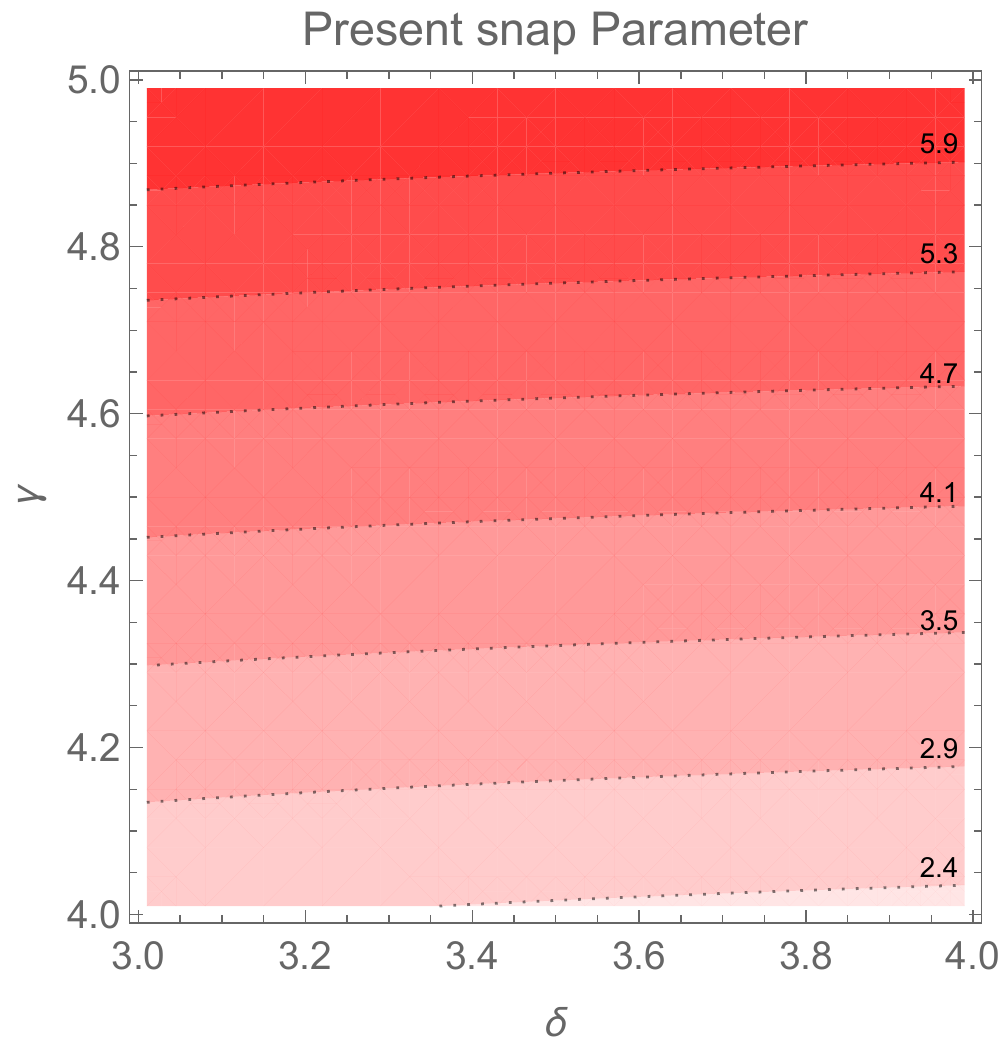}
\caption{Present jerk parameter $j_0$ (left panel) and present snap parameter $s_0$ (right panel) from Eq.\eqref{jerksnap} as a function of $\gamma$ and $\delta$ for the model given in Eq.\eqref{scale1} subjected to the cosmological parameter constraints in Eq.\eqref{cosmotest}. These parameters are shown to vary just slightly with $\delta$ and approximately linearly with $\gamma$.}
\label{fig:jerksnap}
\end{figure*}

This methodology also allows one to obtain a prediction for the cosmological jerk $j$ and snap $s$ parameters (which are sensitive to higher-order derivatives of the scale factor $a\left(t\right)$), which are defined as
\begin{equation}\label{jerksnap}
j=\frac{\dddot a a^2}{\dot a^3},\qquad \qquad  s=\frac{a^{(4)} a^3}{\dot a^4}.
\end{equation}
Similarly as before, we set the values of the exponents $\gamma$ and $\delta$ in the ranges $3<\delta<4<\gamma<5$ and consider the values of the constants $a_s$ and $t_s$ arising from Eq. \eqref{cosmotest}. The predicted present values of $j$ and $s$ can then be obtained by setting $t=t_0$ in Eq. \eqref{jerksnap}. In Fig. \ref{fig:jerksnap} we provide contour plots of the present jerk parameter $j\left(t=t_0\right)\equiv j_0$ and snap parameter $s\left(t=t_0\right)\equiv s_0$ as functions of the exponents $\gamma$ and $\delta$ under the considerations described above. Our results constrain $j$ and $s$ to be in the ranges $j_{\rm min}<j_0<j_{\rm max}$, with $j_{\rm min}\sim2.54$ and $j_{\rm max}\sim 3.54$, and $s_{\rm min}<s_0<s_{\rm max}$, with $s_{\rm min}\sim2.28$ and $s_{\rm max}\sim6.52$. Furthermore, these results indicate that both $j$ and $s$ are only slightly affected by variations of $\delta$ in the range of interest, although they both scale in an approximately linear manner with variations of the exponent $\gamma$. The present parameters $j$ and $s$ have not been measured in experimental data yet. However, a previous work in the generalized hybrid metric-Palatini gravity has shown that a jerk parameter with an order of magnitude $j\sim 1$ would also provide results consistent with the observations of $q_0$ and $H_0$ from the Planck satellite \cite{rosadynsys}.

\section{Conclusions}\label{sec:concl}

In this work, we explored the possibility of sudden singularities arising in the cosmological context of the generalized hybrid metric-Palatini gravity. Similarly to what happens in Brans-Dicke theories of gravity with a parameter $\omega_{BD}=0$, sudden singularities do not arise at the second-order and third-order time derivatives of the scale factor. However, we have shown that, due to the presence of a scalar potential in the modified field equations, sudden singularities are possible in the fourth-order and higher time derivatives of the scale factor, as the second time derivative of the scalar field will contribute with possibly divergent terms proportional to $\ddot\varphi$. For these singularities to appear, one needs the potential $V$ to be at least quadratic in the scalar field $\varphi$, which is associated with the presence of a mass term.

Using an adequate ansatz for the scale factor and a well known form for the potential $V$, which was shown in other works \cite{Rosa:2017jld,Rosa:2018jwp} to considerably simplify  the equations of motion of the scalar fields, we have proposed a broad class of models for which sudden singularities appear in the fourth-order time derivative of the scale factor in FLRW cosmologies. Under an appropriate choice of the normalization constant $V_0$ of the potential, one is able to control the sign of the pressure $p$, thus guaranteeing that the SEC is satisfied throughout the entire time evolution of the system, including at the divergence time $t_s$. Furthermore, we concluded that these results are independent of the spacetime geometry $k$, as the terms dependent on this quantity are always sub-dominant near the sudden singularity.

The cosmological measurements of the Hubble parameter $H_0$, the deceleration parameter $q_0$, and the age of the universe $t_0$ conducted by the Planck Satellite allowed us to impose constraints on the divergence time $t_s$. For our model to be compatible with the experimental measurements of these parameters, the divergence time $t_s$ is constrained to the interval $t_{\rm min}<t_s<t_{\rm max}$, with $t_{\rm min}\sim 3.36t_0$ and $t_{\rm max}\sim 6.56 t_0$. Therefore, the present scenario with sudden singularities cannot be discarded on the grounds of the tests regarding the cosmological background evolution, though the analysis of the perturbations of the current model should be carried out in order to test its viability. In particular, potential deviations from the homogeneity/isotropy assumption of the space-time in the transition from the background cosmological evolution to the perturbations should be addressed in order to examine in detail its effects in the vicinity of the sudden singularity and its consequences for physical observers.

\acknowledgments

JLR was supported by the European Regional Development Fund and the programme Mobilitas Pluss (MOBJD647).
FSNL acknowledges support from the FCT Scientific Employment Stimulus contract with reference
CEECINST/00032/2018 and funding from FCT Projects No. UID/FIS/04434/2020, No. CERN/FIS-PAR/0037/2019 and No. PTDC/FIS-OUT/29048/2017.
DRG is funded by the \emph{Atracci\'on de Talento Investigador} programme of the Comunidad de Madrid (Spain) No. 2018-T1/TIC-10431, and acknowledges further support from the Ministerio de Ciencia, Innovaci\'on y Universidades (Spain) project No. PID2019-108485GB-I00/AEI/10.13039/501100011033, the Spanish project No. FIS2017-84440-C2-1-P (MINECO/FEDER, EU), the project PROMETEO/2020/079 (Generalitat Valenciana), and the Edital 006/2018 PRONEX (FAPESQ-PB/CNPQ, Brazil) Grant No. 0015/2019.



\begin{thebibliography}{99}

\bibitem{Perlmutter:1998np}
S.~Perlmutter \textit{et al.} [Supernova Cosmology Project],
``Measurements of $\Omega$ and $\Lambda$ from 42 high redshift supernovae,''
Astrophys. J. \textbf{517}, 565-586 (1999)
[arXiv:astro-ph/9812133 [astro-ph]].

\bibitem{Riess:1998cb}
A.~G.~Riess \textit{et al.} [Supernova Search Team],
``Observational evidence from supernovae for an accelerating universe and a cosmological constant,''
Astron. J. \textbf{116}, 1009-1038 (1998)
[arXiv:astro-ph/9805201 [astro-ph]].

\bibitem{Copeland:2006wr}
E.~J.~Copeland, M.~Sami and S.~Tsujikawa,
``Dynamics of dark energy,''
Int. J. Mod. Phys. D \textbf{15} (2006), 1753-1936
[arXiv:hep-th/0603057 [hep-th]].

\bibitem{Aghanim:2018eyx}
N.~Aghanim \textit{et al.} [Planck],
``Planck 2018 results. VI. Cosmological parameters,''
Astron. Astrophys. \textbf{641} (2020), A6
[arXiv:1807.06209 [astro-ph.CO]].

\bibitem{Caldwell:1997ii}
R.~R.~Caldwell, R.~Dave and P.~J.~Steinhardt,
``Cosmological imprint of an energy component with general equation of state,''
Phys. Rev. Lett. \textbf{80} (1998), 1582-1585
[arXiv:astro-ph/9708069 [astro-ph]].

\bibitem{Caldwell:2003vq}
R.~R.~Caldwell, M.~Kamionkowski and N.~N.~Weinberg,
``Phantom energy and cosmic doomsday,''
Phys. Rev. Lett. \textbf{91} (2003), 071301
[arXiv:astro-ph/0302506 [astro-ph]].

\bibitem{Ellis:1977pj} 
  G.~F.~R.~Ellis and B.~G.~Schmidt,
  ``Singular space-times,''
  Gen.\ Rel.\ Grav.\  {\bf 8}, 915 (1977).

\bibitem{Nojiri:2005sx}
S.~Nojiri, S.~D.~Odintsov and S.~Tsujikawa,
``Properties of singularities in (phantom) dark energy universe,''
Phys. Rev. D \textbf{71} (2005), 063004
[arXiv:hep-th/0501025 [hep-th]].

\bibitem{Frampton:2011sp}
P.~H.~Frampton, K.~J.~Ludwick and R.~J.~Scherrer,
``The Little Rip,''
Phys. Rev. D \textbf{84} (2011), 063003
[arXiv:1106.4996 [astro-ph.CO]].

\bibitem{Frampton:2011rh}
P.~H.~Frampton, K.~J.~Ludwick, S.~Nojiri, S.~D.~Odintsov and R.~J.~Scherrer,
``Models for Little Rip Dark Energy,''
Phys. Lett. B \textbf{708} (2012), 204-211
[arXiv:1108.0067 [hep-th]].

\bibitem{Dabrowski:2009kg}
M.~P.~Dabrowski and T.~Denkieiwcz,
``Barotropic index w-singularities in cosmology,''
Phys. Rev. D \textbf{79} (2009), 063521
[arXiv:0902.3107 [gr-qc]].

\bibitem{BeltranJimenez:2016dfc}
J.~Beltr\'an Jim\'enez, D.~Rubiera-Garcia, D.~S\'aez-G\'omez and V.~Salzano,
``Cosmological future singularities in interacting dark energy models,''
Phys. Rev. D \textbf{94} (2016) no.12, 123520
[arXiv:1607.06389 [gr-qc]].

\bibitem{Tipler:1977zza}
F.~J.~Tipler,
``Singularities in conformally flat spacetimes,''
Phys. Lett. A \textbf{64} (1977), 8-10.

\bibitem{CK}
C. J. S. Clarke, A. Królak,
``Conditions for the occurence of strong curvature singularities,''
Journal of Geometry and Physics \textbf{2} (1985), 127-143.

\bibitem{Krolak}
A Krolak,
``Towards the proof of the cosmic censorship hypothesis,''
Classical and Quantum Gravity, \textbf{3} (1986), 267.

\bibitem{FernandezJambrina:2006hj}
L.~Fernandez-Jambrina and R.~Lazkoz,
``Classification of cosmological milestones,''
Phys. Rev. D \textbf{74} (2006), 064030
[arXiv:gr-qc/0607073 [gr-qc]].

\bibitem{Jimenez:2016sgs}
J.~Beltran Jimenez, R.~Lazkoz, D.~Saez-Gomez and V.~Salzano,
``Observational constraints on cosmological future singularities,''
Eur. Phys. J. C \textbf{76} (2016) no.11, 631
[arXiv:1602.06211 [gr-qc]].

\bibitem{Clifton:2011jh}
T.~Clifton, P.~G.~Ferreira, A.~Padilla and C.~Skordis,
``Modified Gravity and Cosmology,''
Phys. Rept. \textbf{513} (2012), 1-189
[arXiv:1106.2476 [astro-ph.CO]].

\bibitem{Nojiri:2017ncd}
S.~Nojiri, S.~D.~Odintsov and V.~K.~Oikonomou,
``Modified Gravity Theories on a Nutshell: Inflation, Bounce and Late-time Evolution,''
Phys. Rept. \textbf{692} (2017), 1-104
[arXiv:1705.11098 [gr-qc]].

\bibitem{Saridakis:2021lqd}
E.~N.~Saridakis, R.~Lazkoz, V.~Salzano, P.~Vargas Moniz, S.~Capozziello, J.~B.~Jim\'enez, M.~De Laurentis, G.~J.~Olmo, Y.~Akrami and S.~Bahamonde, \textit{et al.}
``Modified Gravity and Cosmology: An Update by the CANTATA Network,''
[arXiv:2105.12582 [gr-qc]].

\bibitem{Langlois:2017dyl}
D.~Langlois, R.~Saito, D.~Yamauchi and K.~Noui,
``Scalar-tensor theories and modified gravity in the wake of GW170817,''
Phys. Rev. D \textbf{97} (2018) no.6, 061501
[arXiv:1711.07403 [gr-qc]].

\bibitem{Langlois:2018dxi}
D.~Langlois,
``Dark energy and modified gravity in degenerate higher-order scalar\textendash{}tensor (DHOST) theories: A review,''
Int. J. Mod. Phys. D \textbf{28} (2019) no.05, 1942006
[arXiv:1811.06271 [gr-qc]].

\bibitem{Alonso:2016suf}
D.~Alonso, E.~Bellini, P.~G.~Ferreira and M.~Zumalac\'arregui,
``Observational future of cosmological scalar-tensor theories,''
Phys. Rev. D \textbf{95} (2017) no.6, 063502
[arXiv:1610.09290 [astro-ph.CO]].


\bibitem{Barrow:2019cuv}
  J.~D.~Barrow,
  ``Sudden Brans–Dicke singularities,''
  Class.\ Quant.\ Grav.\  {\bf 37}, no. 6, 065014 (2020)
  [arXiv:1909.09519 [gr-qc]].

\bibitem{Barrow:2004xh}
  J.~D.~Barrow,
  ``Sudden future singularities,''
  Class.\ Quant.\ Grav.\  {\bf 21}, L79 (2004)
  [gr-qc/0403084].



\bibitem{Barrow:2004hk}
  J.~D.~Barrow,
  ``More general sudden singularities,''
  Class.\ Quant.\ Grav.\  {\bf 21}, 5619 (2004)
  [gr-qc/0409062].

  \bibitem{Lake:2004fu}
K.~Lake,
``Sudden future singularities in FLRW cosmologies,''
Class. Quant. Grav. \textbf{21} (2004), L129
[arXiv:gr-qc/0407107 [gr-qc]].

  \bibitem{Dabrowski:2005fg}
M.~P.~Dabrowski,
``Statefinders, higher-order energy conditions and sudden future singularities,''
Phys. Lett. B \textbf{625} (2005), 184-188
[arXiv:gr-qc/0505069 [gr-qc]].


\bibitem{Barrow:2020rhh}
  J.~D.~Barrow,
  ``New Anisotropic Sudden Singularities and Dimensional Reduction,''
  Phys.\ Rev.\ D {\bf 102}, no. 2, 024073 (2020)
  [arXiv:2006.14310 [gr-qc]].




\bibitem{Barrow:2004he}
  J.~D.~Barrow and C.~G.~Tsagas,
  ``New isotropic and anisotropic sudden singularities,''
  Class.\ Quant.\ Grav.\  {\bf 22}, 1563 (2005)
  [gr-qc/0411045].

\bibitem{Barrow:2008sn}
  J.~D.~Barrow, A.~B.~Batista, J.~C.~Fabris and S.~Houndjo,
  ``Quantum Particle Production at Sudden Singularities,''
  Phys.\ Rev.\ D {\bf 78}, 123508 (2008)
  [arXiv:0807.4253 [gr-qc]].

\bibitem{Barrow:2009df}
  J.~D.~Barrow and S.~Z.~W.~Lip,
  ``The Classical Stability of Sudden and Big Rip Singularities,''
  Phys.\ Rev.\ D {\bf 80}, 043518 (2009)
  [arXiv:0901.1626 [gr-qc]].

\bibitem{Barrow:2010ij}
  J.~D.~Barrow, S.~Cotsakis and A.~Tsokaros,
  ``The Construction of Sudden Cosmological Singularities,''
  arXiv:1003.1027 [gr-qc].

\bibitem{Barrow:2010wh}
  J.~D.~Barrow, S.~Cotsakis and A.~Tsokaros,
  ``A General Sudden Cosmological Singularity,''
  Class.\ Quant.\ Grav.\  {\bf 27}, 165017 (2010)
  [arXiv:1004.2681 [gr-qc]].

\bibitem{Barrow:2011ub}
  J.~D.~Barrow, A.~B.~Batista, J.~C.~Fabris, M.~J.~S.~Houndjo and G.~Dito,
  ``Sudden singularities survive massive quantum particle production,''
  Phys.\ Rev.\ D {\bf 84}, 123518 (2011)
  [arXiv:1110.1321 [gr-qc]].

\bibitem{Barrow:2013tf}
  J.~D.~Barrow, S.~Cotsakis and A.~Tsokaros,
  ``Series expansions and sudden singularities,''
  arXiv:1301.6523 [gr-qc].

  \bibitem{Perivolaropoulos:2016nhp}
L.~Perivolaropoulos,
``Fate of bound systems through sudden future singularities,''
Phys. Rev. D \textbf{94} (2016) no.12, 124018 [arXiv:1609.08528 [gr-qc]].

\bibitem{Barrow:2013ria}
  J.~D.~Barrow and S.~Cotsakis,
  ``Geodesics at Sudden Singularities,''
  Phys.\ Rev.\ D {\bf 88}, 067301 (2013)
  [arXiv:1307.5005 [gr-qc]].




\bibitem{Barrow:2020ekb}
  J.~D.~Barrow, S.~Cotsakis and D.~Trachilis,
  ``The generic sudden singularity in Brans–Dicke theory,''
  Eur.\ Phys.\ J.\ C {\bf 80}, no. 12, 1197 (2020)
  [arXiv:2009.01732 [gr-qc]].

\bibitem{Tamanini:2013ltp}
N.~Tamanini and C.~G.~Boehmer,
``Generalized hybrid metric-Palatini gravity,''
Phys. Rev. D \textbf{87} (2013) no.8, 084031
[arXiv:1302.2355 [gr-qc]].

\bibitem{Sotiriou:2008rp}
T.~P.~Sotiriou and V.~Faraoni,
``f(R) Theories Of Gravity,''
Rev. Mod. Phys. \textbf{82} (2010), 451-497
[arXiv:0805.1726 [gr-qc]].

\bibitem{Khoury:2003aq}
J.~Khoury and A.~Weltman,
``Chameleon fields: Awaiting surprises for tests of gravity in space,''
Phys. Rev. Lett. \textbf{93} (2004), 171104
[arXiv:astro-ph/0309300 [astro-ph]].

\bibitem{Khoury:2003rn}
J.~Khoury and A.~Weltman,
``Chameleon cosmology,''
Phys. Rev. D \textbf{69} (2004), 044026
[arXiv:astro-ph/0309411 [astro-ph]].

\bibitem{Olmo:2011uz}
G.~J.~Olmo,
``Palatini Approach to Modified Gravity: f(R) Theories and Beyond,''
Int. J. Mod. Phys. D \textbf{20} (2011), 413-462
[arXiv:1101.3864 [gr-qc]].

\bibitem{Gomez:2020rnq}
D.~S\'aez-Chill\'on G\'omez,
``Variational principle and boundary terms in gravity à la Palatini,''
Phys. Lett. B \textbf{814} (2021), 136103
[arXiv:2011.11568 [gr-qc]].

\bibitem{Rosa:2017jld}
J.~L.~Rosa, S.~Carloni, J.~P.~S.~Lemos and F.~S.~N.~Lobo,
``Cosmological solutions in generalized hybrid metric-Palatini gravity,''
Phys. Rev. D \textbf{95} (2017) no.12, 124035
[arXiv:1703.03335 [gr-qc]].

\bibitem{Rosa:2018jwp}
J.~L.~Rosa, J.~P.~S.~Lemos and F.~S.~N.~Lobo,
``Wormholes in generalized hybrid metric-Palatini gravity obeying the matter null energy condition everywhere,''
Phys. Rev. D \textbf{98} (2018) no.6, 064054
[arXiv:1808.08975 [gr-qc]].

\bibitem{Harko:2011nh}
T.~Harko, T.~S.~Koivisto, F.~S.~N.~Lobo and G.~J.~Olmo,
``Metric-Palatini gravity unifying local constraints and late-time cosmic acceleration,''
Phys. Rev. D \textbf{85} (2012), 084016
[arXiv:1110.1049 [gr-qc]].

\bibitem{Capozziello:2012ny}
S.~Capozziello, T.~Harko, T.~S.~Koivisto, F.~S.~N.~Lobo and G.~J.~Olmo,
``Cosmology of hybrid metric-Palatini f(X)-gravity,''
JCAP \textbf{04} (2013), 011
[arXiv:1209.2895 [gr-qc]].

\bibitem{Carloni:2015bua}
S.~Carloni, T.~Koivisto and F.~S.~N.~Lobo,
``Dynamical system analysis of hybrid metric-Palatini cosmologies,''
Phys. Rev. D \textbf{92} (2015) no.6, 064035
[arXiv:1507.04306 [gr-qc]].


\bibitem{Capozziello:2012qt}
S.~Capozziello, T.~Harko, T.~S.~Koivisto, F.~S.~N.~Lobo and G.~J.~Olmo,
``The virial theorem and the dark matter problem in hybrid metric-Palatini gravity,''
JCAP \textbf{07} (2013), 024
[arXiv:1212.5817 [physics.gen-ph]].

\bibitem{Capozziello:2013yha}
S.~Capozziello, T.~Harko, T.~S.~Koivisto, F.~S.~N.~Lobo and G.~J.~Olmo,
``Galactic rotation curves in hybrid metric-Palatini gravity,''
Astropart. Phys. \textbf{50-52} (2013), 65-75
[arXiv:1307.0752 [gr-qc]].

\bibitem{Capozziello:2013uya}
S.~Capozziello, T.~Harko, F.~S.~N.~Lobo and G.~J.~Olmo,
``Hybrid modified gravity unifying local tests, galactic dynamics and late-time cosmic acceleration,''
Int. J. Mod. Phys. D \textbf{22} (2013), 1342006
[arXiv:1305.3756 [gr-qc]].



\bibitem{Capozziello:2013wq}
S.~Capozziello, T.~Harko, T.~S.~Koivisto, F.~S.~N.~Lobo and G.~J.~Olmo,
``Hybrid $f(R)$ theories, local constraints, and cosmic speedup,''
[arXiv:1301.2209 [gr-qc]].

\bibitem{Dyadina:2019dsu}
P.~I.~Dyadina, S.~P.~Labazova and S.~O.~Alexeyev,
``Post-Newtonian Limit of Hybrid Metric-Palatini $f(R)$-Gravity,''
J. Exp. Theor. Phys. \textbf{129} (2019) no.5, 838-848.


\bibitem{Boehmer:2013oxa}
C.~G.~B\"ohmer, F.~S.~N.~Lobo and N.~Tamanini,
``Einstein static Universe in hybrid metric-Palatini gravity,''
Phys. Rev. D \textbf{88} (2013) no.10, 104019
[arXiv:1305.0025 [gr-qc]].


\bibitem{Santos:2016tds}
J.~Santos, M.~J.~Rebou\c{c}as and A.~F.~F.~Teixeira,
``Homogeneous G\"odel-type solutions in hybrid metric-Palatini gravity,''
Eur. Phys. J. C \textbf{78} (2018) no.7, 567
[arXiv:1611.03985 [gr-qc]].


\bibitem{Kausar:2019iwu}
H.~R.~Kausar, R.~Saleem and A.~Ilyas,
``Cosmological inflation in f(X) gravity theory,''
Phys. Dark Univ. \textbf{26} (2019), 100401


\bibitem{Sa:2020qfd}
P.~M.~S\'a,
``Unified description of dark energy and dark matter within the generalized hybrid metric-Palatini theory of gravity,''
Universe \textbf{6} (2020) no.6, 78
[arXiv:2002.09446 [gr-qc]].

\bibitem{Sa:2020fvn}
P.~M.~S\'a,
``Triple unification of inflation, dark energy, and dark matter in two-scalar-field cosmology,''
Phys. Rev. D \textbf{102} (2020) no.10, 103519
[arXiv:2007.07109 [gr-qc]].

\bibitem{Paliathanasis:2020fyp}
A.~Paliathanasis,
``New cosmological solutions in hybrid metric-Palatini gravity from dynamical symmetries,''
[arXiv:2011.05615 [gr-qc]].

\bibitem{rosadynsys}
J. L. Rosa, S. Carloni, J. P. S. Lemos, 
``Cosmological phase space of generalized hybrid metric-Palatini theories of gravity,'' 
Phys. Rev. D \textbf{101}, 104056 (2020).


\bibitem{Lima:2014aza}
N.~A.~Lima,
``Dynamics of Linear Perturbations in the hybrid metric-Palatini gravity,''
Phys. Rev. D \textbf{89} (2014) no.8, 083527
[arXiv:1402.4458 [astro-ph.CO]].

\bibitem{Lima:2015nma}
N.~A.~Lima and V.~S.-Barreto,
``Constraints on Hybrid Metric-palatini Gravity from Background Evolution,''
Astrophys. J. \textbf{818} (2016) no.2, 186
[arXiv:1501.05786 [astro-ph.CO]].


\bibitem{Leanizbarrutia:2017xyd}
I.~Leanizbarrutia, F.~S.~N.~Lobo and D.~Saez-Gomez,
``Crossing SNe Ia and BAO observational constraints with local ones in hybrid metric-Palatini gravity,''
Phys. Rev. D \textbf{95} (2017) no.8, 084046
[arXiv:1701.08980 [gr-qc]].

\bibitem{Avdeev:2020jqo}
N.~Avdeev, P.~Dyadina and S.~Labazova,
``Test of hybrid metric-Palatini f(R)-gravity in binary pulsars,''
J. Exp. Theor. Phys. \textbf{131} (2020) no.4, 537-547
[arXiv:2009.11156 [gr-qc]].

\bibitem{Rosa:2021lhc}
J.~L.~Rosa, F.~S.~N.~Lobo and G.~J.~Olmo,
``Weak-field regime of the generalized hybrid metric-Palatini gravity,''
[arXiv:2104.10890 [gr-qc]].


\bibitem{Koivisto:2013kwa}
T.~S.~Koivisto and N.~Tamanini,
``Ghosts in pure and hybrid formalisms of gravity theories: A unified analysis,''
Phys. Rev. D \textbf{87} (2013) no.10, 104030
[arXiv:1304.3607 [gr-qc]].

\bibitem{Capozziello:2013gza}
S.~Capozziello, T.~Harko, F.~S.~N.~Lobo, G.~J.~Olmo and S.~Vignolo,
``The Cauchy problem in hybrid metric-Palatini f(X)-gravity,''
Int. J. Geom. Meth. Mod. Phys. \textbf{11} (2014) no.5, 1450042
[arXiv:1312.1320 [gr-qc]].


\bibitem{Chen:2020evr}
C.~Y.~Chen, Y.~H.~Kung and P.~Chen,
``Black Hole Perturbations and Quasinormal Modes in Hybrid Metric-Palatini Gravity,''
Phys. Rev. D \textbf{102} (2020), 124033
[arXiv:2010.07202 [gr-qc]].

\bibitem{Kausar:2018ipo}
H.~R.~Kausar,
``Gravitational wave solutions in hybrid metric-Palatini theory,''
Astrophys. Space Sci. \textbf{363} (2018) no.11, 238.

\bibitem{Bombacigno:2019did}
F.~Bombacigno, F.~Moretti and G.~Montani,
``Scalar modes in extended hybrid metric-Palatini gravity: weak field phenomenology,''
Phys. Rev. D \textbf{100} (2019) no.12, 124036
[arXiv:1907.11949 [gr-qc]].

\bibitem{Fu:2016szo}
Q.~M.~Fu, L.~Zhao, B.~M.~Gu, K.~Yang and Y.~X.~Liu,
``Hybrid metric-Palatini brane system,''
Phys. Rev. D \textbf{94} (2016) no.2, 024020
[arXiv:1601.06546 [gr-qc]].

\bibitem{Rosa:2020uli}
J.~L.~Rosa, D.~A.~Ferreira, D.~Bazeia and F.~S.~N.~Lobo,
``Thick brane structures in generalized hybrid metric-Palatini gravity,''
Eur. Phys. J. C \textbf{81} (2021) no.1, 20
[arXiv:2010.10074 [gr-qc]].

\bibitem{Harko:2020oxq}
T.~Harko, F.~S.~N.~Lobo and H.~M.~R.~da Silva,
``Cosmic stringlike objects in hybrid metric-Palatini gravity,''
Phys. Rev. D \textbf{101} (2020) no.12, 124050
[arXiv:2003.09751 [gr-qc]].

\bibitem{Da:2021pbj}
H.~M.~R.~da Silva, T.~Harko, F.~S.~N.~Lobo, J. L. Rosa,
``Cosmic strings in generalized hybrid metric-Palatini gravity,''
[arXiv:2104.12126 [gr-qc]].


\bibitem{Danila:2018xya}
B.~Danila, T.~Harko, F.~S.~N.~Lobo and M.~K.~Mak,
``Spherically symmetric static vacuum solutions in hybrid metric-Palatini gravity,''
Phys. Rev. D \textbf{99} (2019) no.6, 064028
[arXiv:1811.02742 [gr-qc]].

\bibitem{Bronnikov:2019ugl}
K.~A.~Bronnikov,
``Spherically symmetric black holes and wormholes in hybrid metric-Palatini gravity,''
Grav. Cosmol. \textbf{25} (2019), 331-341
[arXiv:1908.02012 [gr-qc]].

\bibitem{Bronnikov:2020vgg}
K.~A.~Bronnikov, S.~V.~Bolokhov and M.~V.~Skvortsova,
``Hybrid metric-Palatini gravity: black holes, wormholes, singularities and instabilities,''
Grav. Cosmol. \textbf{26} (2020) no.3, 212-227
[arXiv:2006.00559 [gr-qc]].

\bibitem{Rosa:2020uoi}
J.~L.~Rosa, J.~P.~S.~Lemos and F.~S.~N.~Lobo,
``Stability of Kerr black holes in generalized hybrid metric-Palatini gravity,''
Phys. Rev. D \textbf{101} (2020), 044055
[arXiv:2003.00090 [gr-qc]].

\bibitem{Danila:2016lqx}
B.~Danila, T.~Harko, F.~S.~N.~Lobo and M.~K.~Mak,
``Hybrid metric-Palatini stars,''
Phys. Rev. D \textbf{95} (2017) no.4, 044031
[arXiv:1608.02783 [gr-qc]].

\bibitem{Capozziello:2012hr}
S.~Capozziello, T.~Harko, T.~S.~Koivisto, F.~S.~N.~Lobo and G.~J.~Olmo,
``Wormholes supported by hybrid metric-Palatini gravity,''
Phys. Rev. D \textbf{86} (2012), 127504
[arXiv:1209.5862 [gr-qc]].

\bibitem{KordZangeneh:2020ixt}
M.~Kord Zangeneh and F.~S.~N.~Lobo,
``Dynamic wormhole geometries in hybrid metric-Palatini gravity,''
[arXiv:2011.01745 [gr-qc]].



\bibitem{Capozziello:2015lza}
S.~Capozziello, T.~Harko, T.~S.~Koivisto, F.~S.~N.~Lobo and G.~J.~Olmo,
``Hybrid metric-Palatini gravity,''
Universe \textbf{1} (2015) no.2, 199-238
[arXiv:1508.04641 [gr-qc]].

\bibitem{Harko:2018ayt}
T.~Harko and F.~S.~N.~Lobo,
``Extensions of f(R) Gravity: Curvature-Matter Couplings and Hybrid Metric-Palatini Theory,''
Cambridge Monographs on Mathematical Physics, Cambridge, Cambridge University Press
(2018).

\bibitem{Harko:2020ibn}
T.~Harko and F.~S.~N.~Lobo,
``Beyond Einstein's General Relativity: Hybrid metric-Palatini gravity and curvature-matter couplings,''
Int. J. Mod. Phys. D \textbf{29} (2020) no.13, 2030008
[arXiv:2007.15345 [gr-qc]].

\bibitem{thesisJL}
J. L. Rosa, ``Cosmological and astrophysical applications of modified theories of gravity'', 
(2019) [arXiv:1911.08257 [gr-qc]].





\end{thebibliography}
\end{document}